# Maximizing spreading influence via measuring influence overlap for social networks

Ning Wang[a], Zi-Yi Wang[b], Jian-Guo Liu[c,1], Jing-Ti Han[c]

[a]*School of Humanities, Shanghai University of Finance and Economics, Shanghai 200433, PR China.*

[b]*School of Information Management and Engineering, Shanghai University of Finance and Economics, Shanghai 200433, PR China.*

[c]*Institute of Fintech, Shanghai University of Finance and Economics, Shanghai 200433, PR China.*

**Abstract:**

Influence overlap is a universal phenomenon in influence spreading for social networks. In this paper, we argue that the redundant influence generated by influence overlap cause negative effect for maximizing spreading influence. Firstly, we present a theoretical method to calculate the influence overlap and record the redundant influence. Then in term of eliminating redundant influence, we present two algorithms, namely, Degree-Redundant-Influence (DRS) and Degree-Second-Neighborhood (DSN) for multiple spreaders identification. The experiments for four empirical social networks successfully verify the methods, and the spreaders selected by the DSN algorithm show smaller degree and *k*-core values.

**Introduction:**

Maximizing influence spreading for social networks is implemented by identifying a set of influential nodes as initial spreaders to active maximized excepted number of nodes, which has attracted much attention in recent years (Kempe et al., 2003a; Cao et al., 2011; Morone and Makse, 2015). For the multiple spreaders identification, the interactions among spreaders play more important role compared with single vital node identification, regarding to structure properties and activity patterns among the spreaders (Lü et al., 2016). As the influential nodes in social networks generally tend to be concentrated (Zhou et al., 2017), the influence thet spread will overlap with others to generate redundant influence which is invalid for influence maximization (Jung et al., 2012; Zhou et al., 2018). As shown in Fig.1, the total influence exerted on node $v_1$ from spreaders $s_1$ $s_2$ and $s_3$ has exceed its upper limit and generate redundant influence from the view of infected probability. Recent researches have paid attention to reducing influence overlap, including decreasing spreaders similarity (Liu et al., 2017), keeping spreaders disconnected (Ma et al., 2016) and iteratively removing the selected spreaders (Chen and Wang, 2009) to maximize spreading infleunce. However, only excessive influence overlap will cause nagetive effect for influence maximization. Empirical analysis found that information can diffuse to the whole network if it breaks through the partial blockade (Hu et al., 2018). And the influence overlap of spreaders in local region can produce synergistic effects to improve information spreading ability (Flores et al., 2012;

---

[1] Corresponding author.

*Email address:* liujg004@ustc.edu.cn (Jian-Guo Liu )
Tel: +86 152 0212 1269

Piedrahita et al., 2018). As showed in Fig.1, the influence overlap of spreaders $s_1$ and $s_2$ has increased the infected probability of node $v_3$ while not cause redundant influence. So eliminating the redundant influence generated by the influence overlap is one of the key issue for maximizing spreading influence, rather than completely reducing influence overlap. While previous work focused on reducing the influence overlap of spreaders and few researches have provided metrics to measure the part of redundant influence generated by influence overlap. In this paper, we identify the redundant influence from influence overlap and argue its negative effect for maximizing spreading influence.

Firstly, the influence overlap is defined as an intuitive phenomenon that existing more than one spreaders can influence the same non-spreader node together (Granovetter, 1977; Yang and Huang, 2017), which is showed in Fig.1(a). Previous work have quantified the extent of overlap directly from the perspective of spreaders by their similarity (Liu et al., 2017), cosine (Sohn, 2001) and Euclidean distance (Burt and Talmud, 1993), but these methods failed to calculate the redundant influence generated by the influence overlap. To solve this problem, we present a theoretical method to measure the overlap influence from the view of non-spreader nodes based on their infected probability, and quantify the influence overlap by calculating the total influence exerted on each non-spreader node. The redundant influence exerted on a non-spreader node $v$ as shown in Fig.1(b), is defined as the part of total influence exceeding its upper limit, which is denoted as RI($v$).

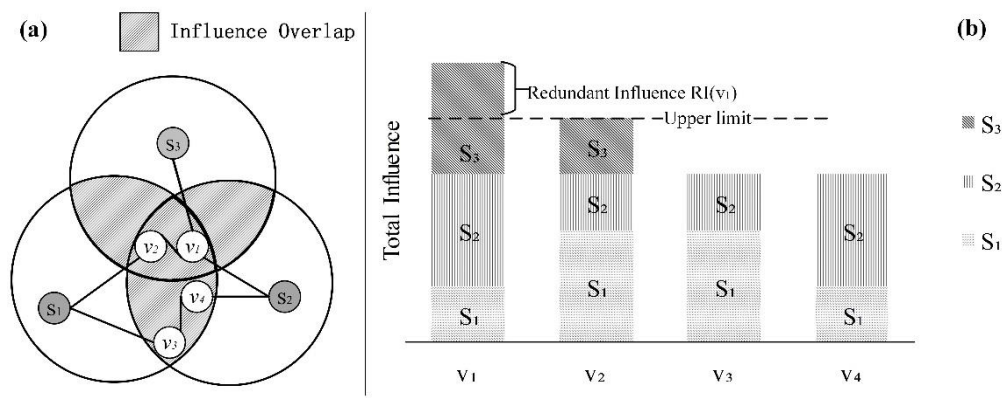

**Figure 1.** Illustration of the *Influence Overlap* (a) and *Redundant Influence* (b).

Secondly, we present two algorithms for identifying multiple spreaders in term of eliminating the Redundant Influence to verify the validity of our method. One is a iterative algorithm namely Degree-Redundant-Influence (DRI) and the other one is a heuristic algorithm namely Degree-Second-order-Neighborhood (DSN). The DRI method constructed by eliminating the redundant influence when iteratively identifying each spreader. While the DSN method identifies the spreaders by a special rule resulting from an optimal problem aiming at minimizing redundant influence.

**Literature Review**

Directly finding out the spreader set to achieve maximized spreading influence has been proved to be an NP-hard problem (Kempe et al., 2003), so lots of algorithms have been proposed to select the most influential nodes to consist the spreader set (Ott et al., 2018; Valente and Fujimoto, 2010). One way is directly selecting spreaders according to their importance ordered by various centralized indices. Some based on local structure information such as degree centrality (Pastor-Satorras and Vespignani, 2002), nodes clustering coefficient (Ertem et al., 2016a), and others utilize the whole network structure, such as

*k*-core decomposition (Larsen and Ellersgaard, 2017), *h*-index (Lü et al., 2016b), node betweenness (Bozzo and Franceschet, 2013). While purely utilizing these indices are not enough to distinguish the importance of each node, for example, Lin (Lin et al., 2014) found that the spreading influence for different nodes of the largest *k*-core node set are different, and Korn (Korn and Muthukrishnan, 2000) have proved that the neighbors of a node can also affect its influence spreading ability. By considering the neighbors' *k*-core value, Bae (Bae and Kim, 2014) offered a two-step iterative algorithm, namely Neighborhood Coreness based method (NC), to integrate both local and whole network structure information. Although these methods are valid to distinguish the importance of different nodes, they ignored the phenomenon of influence overlap among spreaders (Chang et al., 2018).

The characteristic of influence dynamics and social network structure determine that influence overlap is important for influence maximization (Lü et al., 2016a). Empirical studies found that the structure of social network often present first- or second-order positive assortativity coefficient (Wood, 2017; Zhou et al., 2017), that is, the influential spreaders often connect with each other or have common friends. Therefore, identifying spreaders only based on centralized indices will makes the influence of one spreader easily to overlap with others, which may further generate redundant influence (Z.-K. Zhang et al., 2016). Recent studies have considered the effect of influence overlap and provided some algorithms for identifying multiple spreaders. One way is increasing the distance between spreaders to decrease the overlap influence area (Guo et al., 2016; Hu et al., 2014), the other one is removing the selected spreaders for each spreader identification (Chen and Wang, 2009; J. X. Zhang et al., 2016).

When using the distance to reduce the influence overlap, mainly use the Euclidean distance and local similarity between the nodes. Kitsak (Kitsak et al., 2010) found in their experiment that keeping spreaders disconnected can largely improve the influence spreading for multiple spreaders identification, no matter selecting spreaders based on node degree or *k*-core values. They demonstrated the importance of distance parameter among spreaders, which could effectively reduce influence overlap, but whether the distance is powerful in different situations and what is the optimal distance among spreaders are still unclear in their paper. Hu (Hu et al., 2014) had investigate the relationship of spreaders' distance and influence spreading, the result showed that the effective of distance is affected by the network types, and in the same network, larger average degree and connection probability will result in a smaller optimal distance of the most effective spreading. And Guo (2016) demonstrated that adding distance parameter into coloring method (Zhao et al., 2014) can also obviously improve influence spreading, where each node is colored and the distance between initial nodes is equal to average network distance. As finding the optimal distance among each pair of spreaders is difficult, Liu (Liu et al., 2017) proposed local structure similarity (LSS) algorithm to reduce influence overlap by iteratively selecting the spreader whose similarity is below the artificial similarity threshold *r* with each selected ones. All above methods deal with the effect of influence overlap, but how to find the optimal parameters (i.e. spreaders distance, similarity threshold r) in different networks is still unclear.

The second way to reduce influence overlap is iteratively removing the selected spreaders in the network. Chen (Chen and Wang, 2009) proposed a Degree Discount (DD) algorithm that in each round, selecting the node based on degree centrality in the network and removing the selected spreader and its edges. Morone (Morone and Makse, 2015) firstly mapped the influence maximization problem into percolation theory to break the connection of network. They proposed a Collective Influence (CI) greedy algorithm, which determines the importance of the node via the degree of neighbors. And during each round of spreaders identification, they select the most important one and remove it from the network. Besides directly removing the selected spreaders, Zhang (J. X. Zhang et al., 2016) presented an iterative method called "VoteRank", where all nodes vote in a spreader in each turn, and the voting ability of neighbors of elected spreader will be decreased in subsequent turn.

All these multiple spreaders identification algorithms focused on the development of selection rules to reduce spreaders' influence overlap, regardless maximizing the spreading influence in terms of the influence overlap. An interesting phenomenon observed by Hu (2018) is that the influence spreading often exhibits double peaks, that is, only by infecting enough local nodes can the influence be disseminated to the entire network. The empirical study about obesity spreading showed that the influence of a single people is within his/her three-order neighbors and decays exponentially with the propagation distance (Christakis and Fowler, 2007). So the overlap of spreaders' influence can improve the infected probability of the nodes located at their common second or third-order neighbors, which is helpful for the influence diffused to the whole network. The influence overlap is therefor like a double-edged sword, proper overlap can improve the influence spreading, but excessive overlap will create negative effect (Liu et al., 2018). Thence, quantifying the influence overlap and eliminating the corresponding redundant influence play an crucial role for maximizing spreading influence. Inspired by this idea, we present a theoretical method to calculate the influence overlap and record the redundant influence exerted on each non-spreader nodes, which is denoted as RI(*v*). By eliminating the redundant influence calculated by our method, we present a greedy algorithm (DRI) and a heuristic algorithm (DSN) to compare with the state-in-the-art methods on SIR information spreading model.

The reminder of the paper is organized as follows: We will present a theoretical method to quantify the influence overlap and record the redundant influence, and present two methods based on eliminating redundant influence in Section 2. In Section 3, we will describes the data used in this paper. Section 4 and 5 respectively shows the experiment results analysis and discusses the practical contribution of our algorithms. Finally, the conclusions and directions for the future work are given in Section 6.

**2. Methodology**

2.1 The definition of influence overlap

Influence overlap used to be described as a phenomenon in previous work, that is, more than two spreaders can infect the same node (Liu et al., 2017) . But the methods measuring the influence overlap habitually from the perspective of the spreaders, failed to quantify the redundant influence exerted on each non-spreader nodes (Sohn, 2001 & Burt and Talmud, 1993). So, in this paper, we give mathematical expressions of influence overlap as $\exists s_i, s_j$ fit *I(v,$s_i$)>0* and *I(v,$s_j$)>0* where $s_i, s_j$ denotes any two spreaders and *I(v, s)* denotes the influence of spreader *s* exert on the node *u*. And the total influence exerted on node *v* from spreader set $\{s_i, s_j\}$ is denoted as $I(v, \{s_i, s_j\})$.

2.2 Quantifying the redundant influence

For a given network *G* ={*V,E*} where *V* is the set of nodes and *E* is the set of links in the network. Let $S = \{s_1, s_2, ..., s_m\}$ represent the *m* spreaders in the network. We use *I(v,s)* to measure the influence of spreader *s* exerted on node *v*, which is usually expressed by the probability that node *v* could be infected by spreader *s* in SIR model (Fang and Hu, 2018; Morales et al., 2014; Sewell, 2018). As the probability that node *v* will be infected cannot excess 1, so we have the upper limit of $I(v, s)$ fit:

$$I(v,s) \leq 1 . \tag{1}$$

If we consider the spreader set $S = \{s_1, s_2, ..., s_m\}$ as a whole to analyze the probability that they can infect a node $v$, we can get the upper limit of the probability that node $v$ would be infected by the spreader set $S$ as follows:

$$I(v,S) \leq 1 \text{ for } v \in V - S . \tag{2}$$

After obtaining the upper limit of the infection probability of each node, we first calculate the probability of infection between any pair of nodes. Empirical studies (Christakis and Fowler, 2007) found that the valid influence range of a single person is within his/her third-order neighbors, and the strength of influence is generally exponentially decreasing as the range of propagation increases, which is showed at Fig.2(a) and denoted as

$$I(v,s_i)_{s_i \in N_{1(v)}} \approx I(v,s_j)^2_{s_j \in N_{2(v)}} \approx I(v,s_p)^3_{s_p \in N_{3(v)}} , \tag{3}$$

where $N_{i(v)}$ represents the set of $i_{th}$-order neighbors of node $v$. The Eq. (3) indicates that if a node $v$ is infected by one of his/her nearest friend with probability $\beta$, then it will be infected by one of his/her second-order friend with probability $\beta^2$, and be infected by one of his/her third-order friend with probability $\beta^3$. The decreasing relationship of influence spreading is consistent with the attenuation process when information is spreading on a single propagation path. So the infected probability between two nodes can decided by their relative location, which is determined by their shortest propagation path as showed in Fig.2.

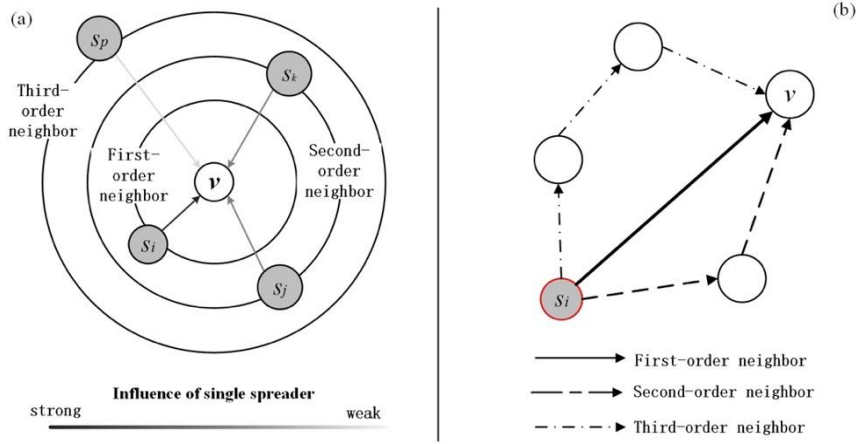

Figure2: Illustration of spreader influence attenuation (a) and the definition of relative position (b)

Then we decompose the total influence exerted on a non-spreader node $v$ based on the relative positions with each spreader. As the valid influence range of spreaders is limited within his/her third-order neighbors (Christakis and Fowler, 2007), we can substitute the $I(v, S)$ as followed:

$$I(v,S) = \sum_{i=1}^{3} \sum_{S_{iv}=(S \cap N_{i(v)})} I(v, S_{iv}) , \tag{4}$$

where $S_{iv} = (S \cap N_{i(v)}) = \{s_{iv}^1, s_{iv}^2, ... s_{iv}^{n_i}\}$ denotes the part of the spreader set $S$ that locate at the $i_{th}$-order neighbors of a non-spreader node $v$, and $n_i$ represent the number of spreaders in the $S_{iv}$. In this paper, as we assumed that the spreading process starting simultaneously, so the spreaders locating at different order neighbors of node *v* will affect it at different time. Hence, the total influence exerted on the node *v* can be divided into three parts from different order neighbors (Zhou et al., 2018).

Next, we analyze the spreaders in the same order neighbors of node *v*, such as $s_j$ and $s_k$ in the Fig.2 (a), which can influence the non-spreader node *v* at the same time. In the SIR model, the influence from different spreaders are independent from each other, so the probability of node *v* will be infected by spreaders $s_j$ and $s_k$ follows Binomial distribution. In this paper, for a given infected probability $\beta$, the probability of node *v* be infected by the spreaders $s_j$ and $s_k$ in the Fig.2 is denoted as:

$$I(v, \{s_j, s_k\} \mid \beta) = 1 - (1 - \beta^2)^2, \tag{5}$$

where $\beta^2$ is the probability for the node *v* will be infected by the spreader $s_j$ and $s_k$ which locate at the second-order neighbors of node *v*. Therefore, we abstract the influence from the part of spreader set $S_{iv}$ which locate at $i_{th}$-order neighbors of node *v* as followed:

$$I(v, S_{iv} \mid \beta) = I(v, n_i \mid \beta) = 1 - (1 - \beta^i)^{n_i}, \tag{6}$$

where $n_i$ is the number of spreaders of $S_{iv}$. Combining Eq. (4) and Eq. (6), we can express the total influence from spreader set *S* exerted on arbitrary non-spreader node *v* *under a given infected probability $\beta$* as followed:

$$I(v, n_1, n_2, n_3 \mid \beta) = [1 - (1 - \beta^1)^{n_1}] + [1 - (1 - \beta^2)^{n_2}] + [1 - (1 - \beta^3)^{n_3}]. \tag{7}$$

Finally, based on Eq.(1) and Eq.(7), we can calculate the total influence exerted on a non-spreader node *v* from the spreader set *S* under the infected probability $\beta$. The Redundant Influence of a non-spreader node *v* is the part of total influence exceeding its upper limit which is showed as followed:

$$RI(v, n_1, n_2, n_3 \mid \beta) = \begin{cases} 0 & when\ I(v, n_1, n_2, n_3 \mid \beta) \leq 1 \\ I(v, n_1, n_2, n_3 \mid \beta) - 1 & when\ I(v, n_1, n_2, n_3 \mid \beta) > 1 \end{cases}. \tag{8}$$

The Ep.(8) indicates that the influence overlap will cause negative effect when existing redundant influence, which is affected by both the infected probability and the location of spreaders. Then to verify the validity of this theoretical method, we present two algorithms for multiple spreaders identification based on eliminating the redundant influence calculated by Eq.(8).

2.2 The Degree-Redundant-Influence algorithm

Eliminating the redundant influence is important for influence maximization, but it is not reasonable to select all the spreaders at the marginal of the network, so we have to make trade-offs between selecting influential nodes and eliminating redundant influence. Lots of indices have been proposed to characterize influential nodes (Harrigan et al., 2012; Larsen and Ellersgaard, 2017), while considering the simplicity and practicality of the operation, we choose the node degree as the index for measuring node importance. Combining the node degree index and the method for calculating redundant influence, we present a new algorithm for multiple spreaders identification, namely Degree-Redundant-Influence algorithm (DRI). The operation steps of the DRI algorithm shows as followed, and the core idea of DRI algorithm is that keeping no redundant influence for all the non-spreader nodes when iteratively selecting each spreaders. In addition, to improve the efficiency of DRI algorithm, we filter out the nodes with the degree 1, as these marginal nodes are lack of spreading ability. Compared with the previous greedy algorithms (Xia et al., 2016) that converge when selecting fix number of spreaders $m$, the DRI can also converge automatically when

---

The DRI algorithm

Input: Network $G=(V,E)$, infect probability $\beta$
Output: Spreader set $S \leftarrow \{s_1, s_2, \ldots, s_m\}$

1. Sort nodes by degree $V' \leftarrow \{v'_1, v'_2, \ldots v'_N\}$ where $deg(v'_1) \geq deg(v'_2) \geq \cdots \geq deg(v'_N)$
2. $S \leftarrow \emptyset$
3. **for** $i \leftarrow 1\ to\ n$ **do**
4.     **if** len(S) >$m$ **then**
5.         exit                    # converge condition 1.
6.     **else if** $deg(v'_i)$ >1 **then**     # filter out the marginal nodes
7.         $S \leftarrow S \cup v'_i$            # take node $v'_i$ as an alternative spreader
8.         **for** $u_j \in V$-$S$ **do**
9.             $RI \leftarrow RI(u_j)$       # use Eq.(8) to calculate redundant influence
10.            **if** $RI > 0$ **then**     # exist node $u_j$ which has redundant influence
11.                S$\leftarrow S - v'_i$     # $v'_i$ cannot be a spreader and try $v'_{i+1}$
12.                exit
14.     **end**

---

arbitrarily adding a spreader will have redundant influence, which is helpful for managers to quickly figure out how many spreaders they real need in practice application.

| | | |
|---|---|---|
| 15. | **else** | |
| 16. | **exit** | # converge condition 2. |
| 18. | **end** | |
| 19. | **return** *S* | |

2.3 The Degree-Second-order-Neighbors algorithm

    The DRI algorithm can completely eliminate the redundant influence, while traversing the network continuously to judge where existing redundant influence for all non-spreader nodes cost so much time. So for large-scale social networks, we present a heuristic algorithm to decrease computing complexity for spreaders identification. By analyzing Eq.(7) and Eq.(8), we find that the redundant influence is affected by the number of spreaders located within the third-order neighbors of a non-spreader node *v* ($n_1, n_2, n_3$) and the infected probability ($\beta$) between nodes. In this paper, we assume that the network is homogeneous, that is, the probability of infection between nodes is the same. Therefore, the only way to eliminate redundant influence is changing the location of spreaders. Inspired by this idea, we create an optimal question to find the feature of spreaders location targeting on keeping no redundant influence.

    The target function is to find the maximizing number of spreaders that located within three-order

Table 1: The solutions of the optimal problem with different $\beta$,

neighbors of a non-spreader node without redundant influence. The varibles $x_1, x_2, x_3$ represent the number of spreaders respectively located at the first-order, second-order and third-order neighbors of a non-spreader node *v*. At the same time, we set that $x_1 \leq x_2 \leq x_3$, because from the perspective of universality, the number of high-order neighbors is much larger than the number of nodes in low-order neighbors, so there will be more spreaders in high-order neighbors. On the other hand, this condition also limits the concentrate location of spreaders around a certain node. Therefore, under a fixed infected probability $\beta$, for an any non-spreader node *v*, the optimal problem can be denoted as followed:

$$\text{Maximize} \quad \text{argmin RI}(v, x_1, x_2, x_3 \mid \beta) \qquad (9)$$
$$\text{Subject to} \quad \begin{cases} x_1 \leq x_2 \leq x_3 \\ x_1, x_2, x_3 \in N \\ \beta \in R^+ \end{cases}$$

    Solving the above optimal problem, we can get different solutions of $(x_1, x_2, x_3)$ for different value of infected probability $\beta$, the results are presented in Table1. From the Table1, we find that the value of $x_1$ is the dominating factor to cause redundant influence, for example, increasing the value of $x_1$ will significantly decrease the maximum value of $x_2$ and $x_3$ (i.e: for $\beta$=0.3, $(x_1, x_2, x_3) = (1,7,8)$ or $(2,4,7)$). The most important finding is that when there are two or more spreaders in the first-order neighbors of the non-spreader node *v*, it is more likely to cause redundant influence on node *v*. In other words, the first-order neighbors shared by spreaders are prone to have redundant influence, such as the node $v_1$ in Fig.1. Thus in order to reduce the redundant influence, there is preferably no common first-order neighbors between any pair of spreaders, which means the distance between spreaders is preferably greater than two. This finding is also consistent with lots of previous studies (Liu et al., 2016; Shrestha et al., 2015; Yang and Huang, 2017) that measuring the influence of a spreader within its second-order neighbors.

|  | Maximize $(x_1, x_2, x_3)$ | | |
|---|---|---|---|
|  | $x_1$ | $x_2$ | $x_3$ |
| $\beta = 0.50$ | 1 | 1 | 2 |
| $\beta = 0.50$ | 0 | 3 | 4 |
| $\beta = 0.45$ | 1 | 2 | 2 |
| $\beta = 0.45$ | 0 | 4 | 5 |
| $\beta = 0.40$ | 1 | 2 | 6 |
| $\beta = 0.40$ | 0 | 6 | 6 |
| $\beta = 0.35$ | 2 | 2 | 4 |
| $\beta = 0.35$ | 1 | 4 | 6 |
| $\beta = 0.35$ | 0 | 8 | 9 |
| $\beta = 0.30$ | 3 | 3 | 3 |
| $\beta = 0.30$ | 2 | 4 | 7 |
| $\beta = 0.30$ | 1 | 7 | 8 |
| $\beta = 0.30$ | 0 | 12 | 13 |
| $\beta = 0.25$ | 4 | 4 | 5 |
| $\beta = 0.25$ | 3 | 5 | 6 |
| $\beta = 0.25$ | 2 | 8 | 9 |
| $\beta = 0.25$ | 1 | 12 | 13 |
| $\beta = 0.25$ | 0 | 20 | 20 |

The DSN algorithm

Input: Network $G=(V,E)$, infect probability $\beta$, the number of the spreaders m
Output: Spreader set $S = \{s_1, s_2, ..., s_m\}$

1. Sort nodes by degree $V' \leftarrow \{v'_1, v'_2, ... v'_N\}$ where $deg(v'_1) \geq deg(v'_2) \geq \cdots \geq deg(v'_N)$
2. $S \leftarrow \emptyset$
3. **for** $i \leftarrow 1\ to\ n$ **do**
4.     **if** len(S) >m **then**
5.         **exit**                        # the converge condition
6.     **else if** $deg(v'_i) > 1$ **then**      # filter out the marginal nodes
7.         $sign \leftarrow 1$                # record whether the alternative spreader $v'_i$ fit the condition

| | | |
|---|---|---|
| 8. | **for** $s_i \in S$ **do** | |
| 9. | $\quad Dis \leftarrow \text{Distance}(v_i', s_i)$ | # calculate the distance between node $v_i'$ and spreaders |
| 10. | $\quad$ **if** $Dis < 3$ **do** | # the identifying condition |
| 11. | $\quad\quad sign \leftarrow 0$ | |
| 12. | $\quad\quad$ **exit** | |
| 13. | **end** | |
| 14. | **if** $sign = 1$ **then** | |
| 15. | $\quad S \leftarrow S \cup v_i'$ | # sign=1 indicates the node $v_i'$ fit the identifying conditions |
| 16. | **end** | |
| 17. | **Return** $S$ | |

In term of this important finding, we present a new heuristic algorithm, namely Degree-Second-order-Neighbor algorithm (DSN), for multiple spreaders identification. We still utilize the node degree to represent node importance and request the distance between each pair of spreaders should greater than two to eliminate redundant influence. The detail process of the DSN method is showed in following table. The core step of DSN algorithm is judging whether the shortest distance between the alternative node and all the selected spreaders exceed two. To improve the efficiency of the DSN algorithm, we also filter out the marginal nodes in the network as DRI algorithm. The DSN method is suitable for large-scale social network for avoid repeating scan the whole network, and the running time of DSN algorithm is between $O(\log n + m(m+1)/2)$ to $O(\log n + mn)$ corresponding to the best and worst scenarios. The $m$, $n$ are the number of spreaders and total nodes in the network.

### 3. Data description

The empirical analysis is based on four online social networks with different network topologies which can significantly affect the influence spreading (Piedrahita et al., 2018), including node average degree (Harrigan et al., 2012), clustering coefficient (Ventresca and Aleman, 2013; Vriens and Corten, 2018) and average nodes distance (Yamaguchi, 2002). We have choose four empirical datasets with different social relationship and structural characteristics from openly accessed (Rossi and Ahmed, 2015), including Ca-GrQc, email, Soc-hamsterster and Facebook networks.

The Ca-GrQc network (Rossi and Ahmed, 2015) is an Arxiv General Relativity and Quantum Cosmology collaborate network in which nodes and links represent scientists and scientific collaborations. Such networks tend to have local clustering structure and a little connections between different communities, as establishing scientific collaborations is slow and difficult (De Stefano et al., 2013; Lopaciuk-Gonczaryk, 2016). For example, scholars in the same laboratory will cooperate with each other, and scholars' exchanges between different institutions will connect some small groups together. Therefore, the nodes in Ca-GrQc network have higher clustering coefficient, small average degree and long average distance.

The email network (Guimera et al., 2003; Rossi and Ahmed, 2015) is collected from the University Rovirai Virgili in Tarragona in the south of Catalonia in Spain. Nodes in this email communication network represents the users, and the edges indicate that at least one email has sent between the users. Email network is an important medium for information dissemination(Johnson et al., 2012), for example merchants often use email to recommend their products to users, and good products will further put forwarded to their friends by these users. The establishment of email relationships is much easier than scientific collaboration, so compared with the Ca-GrQc networks, the average degree of each nodes will be bigger and the average nodes distance will be shorter. At the same time, most people in the email network are not familiar with each other, so the nodes clustering coefficient will be relatively lower in email network.

The Soc-hamsterster (Hamsterster; Rossi and Ahmed, 2015b) network consist of the friendships and family links between the users of the website, where people can share the information about the movies, music and so on. Compared with scientist network and email network, it is easier to establish relationship based on common hobbies, for example, people can freely exchange their opinion about the movies and music with any strangers without any cost. So the nodes in the network will have more average friends and tend to be close with each other.

The Facebook social network (Traud et al., 2012; Rossi and Ahmed, 2015b) is an online social friendship network extracted form Facebook. Apparently, the users are the nodes and online friendship ties are the edges in this network. The edges are considering without direction and weight, i.e. an edge exits as long as user pages reciprocate "friendship" between them. We apply the largest connected component of the network in this work since we would test spreading influence in the connected network.

Four network datasets employed in this work are all undirected and unweighted. The descriptive statistics of each network are shown in the Table 2. The network size refer to the number of nodes $N$ and edges $E$. The average degree $\langle k \rangle$ show the information of density of the network. The average distance $\langle d \rangle$ offers the average shortest number of steps from one node to another. The clustering coefficient $C$ represent the ratio that the friends of the same node connect with each other. Additionally, these four empirical networks illustrate the highly heterogeneous nature of scale-free networks, so we calculated their respective epidemic thresholds according to the method $\beta_c = \langle k \rangle / \langle k^2 \rangle$ (Pastor-Satorras and Vespignani, 2002) and used $\beta_c$ as the basis for selecting $\beta$ in our experiments.

Table 2: The descriptive statistics of the Ca-GrQc, Email, Soc-hamsterster and Facebook networks, including the number of nodes $N$, the number of edges $E$, the average degree $\langle k \rangle$, the average distance between each pair of nodes $\langle d \rangle$ and clustering coefficient $C$ and the epidemic thresholds $\beta_c$.

| Network | $N$ | $E$ | $\langle k \rangle$ | $\langle d \rangle$ | $C$ | $\beta_c$ |
|---|---|---|---|---|---|---|
| Ca-GrQc | 4,158 | 13,422 | 6.46 | 6.05 | 0.56 | 0.06 |
| Email | 1,133 | 5,450 | 9.62 | 3.61 | 0.22 | 0.05 |
| Soc-hamsterster | 2,000 | 16,097 | 16.09 | 3.59 | 0.54 | 0.02 |
| Facebook | 30,106 | 1,176,489 | 78.16 | 3.06 | 0.21 | 0.006 |

## 4. Experimental result analysis

In this part of work, we compare our two algorithms DRI and DSN with three state-in-the-art algorithms to verify the negative effect of redundant influence for the influence maximization. We choose the Neighborhood Coreness based algorithm─NC (Bae and Kim, 2014) as a baseline algorithm as it ignores the effect of influence overlap when identifying multiple spreaders. At the same time, we compare our algorithms with the Node Disconnect algorithm─ND (Kitsak et al., 2010) and Collective Influence algorithm─CI (Morone and Makse, 2015), which have consider the effect from influence overlap in two different views, but failed to distinguish the redundant influence. The main experimental process consists of two steps. First, we identify the same number of spreaders by different algorithms for each network. It is worth pointing out that the DRI algorithm has two convergence conditions, so the number of spreaders on some networks will be smaller than other algorithms. Then, we implement the simulation under SIR diffusion model and utilize the average infected fraction (AIF) to express the spreading influence. Based on the simulation results, we further analyze the relationship between the redundant influence and the maximum influence produced different spreaders. The results show that with the same infected probability $\beta$, when increasing the number of spreaders *m*, the emergence of redundant influence is accompanied by the convergence of the maximum influence.

All the experiments conducted on a computer running an Intel i7-2600k processor at 3.4GHz with 16GB of RAM, and the time spent on networks generation is lower than 5 second even size of network is over 30,000 nodes. Each simulation repeated for 100 times to ensure the stability of experimental results.

4.1 SIR diffusion model

The SIR model was initially used to describe the mechanism of disease transmission and was later widely used in various information dissemination models (Iribarren and Moro, 2011; Ventresca and Aleman, 2013).The nodes in SIR model have three different states: susceptible, infected and recover. The infected nodes can infect susceptible one that change its state into infected, but the recover one will not change its state again. In the simulation, the influence spreading is step by step, and in each step, the infected nodes will infect their susceptible friends in probability $\beta$, then transform to the state of recover in probability $\gamma$. From the perspective of influence maximization, it will affected by the ratio of $\beta/\gamma$ and number of spreaders *m*. So we assume that the recovery probability of all infected nodes is 1and only change the value of $\beta$ as previous work (Wang et al., 2017; Zhou et al., 2018). Besides this, we assume that the network is homogeneous and so that the infected probability $\beta$ is the same between all the nodes.

4.2 Result analysis

We compare these five algorithms based on the average infected fraction (AIF) on four empirical datasets. In this paper, we set the number of spreaders in each network range from 10 to 100 and separated by10, and we set the infected probability $\beta$ in Ca-GrQc network, Email network and Soc-hamsterster network changing from 0.2 to 0.4 and separated by 0.05, the $\beta$ in Facebook network changing from 0.03 to 0.09 and separated by 0.02. The setting of $\beta$ is in term of the epidemical threshold $\beta_c$ of the corresponding networks. We show the change of AIF with spreading steps (T) with a specifically pair of $m$ and $n$ in Fig. 3, where $m$=100 in four social networks except $m_{DRI}$=30 in Email network and $m_{DRI}$=48 in Soc-hamsterster network. From Fig.3, one can find that the AIF produced by the spreaders identified by DSN algorithm is larger than other algorithms on Ca-GrQc, Email and Soc-hamsterster networks. Additionally, the number of spreaders identified by DRI algorithm is less than ND, CI and NC algorithms, but produce more AIF than the spreaders identified by these three methods. As for ignoring the effect of influence overlap, the spreaders identified by NC algorithm can diffuse the influence faster in the early stage of influence propagation but produce less AIF when spreading convergence. We also find an interesting phenomenon that all the spreaders have produced similar AIF on Facebook network, where the spreading influence have diffused to the most of the nodes in the network and convergent fast within four steps. The simulations of other pair of $m$ and $\beta$ have the same result and we will show them in Appendix.

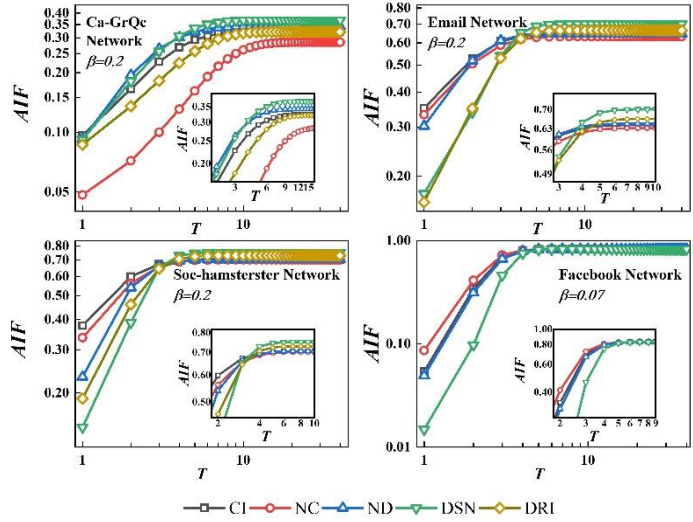

Figure 3: (Color online) The average infected fraction (AIF) with the spreading step (T). Each algorithm selects 100 spreaders in each networks, except DRI algorithm only identifies 48 spreaders for Soc-hamsterster network and 30 spreaders for email network. From figure one can find that the spreaders selected by DSN algorithm produce larger AIF on Ca-GrQc, Email and Soc-hamsterster networks.

Based on the simulation results, we calculated the Redundant Influence (RI) generated by the spreaders on CA-GrQc network, Email network and Soc-hamsterster network, and analyze the relationship between the RI and corresponding AIF. As the spreaders identified by the DRI algorithm naturally eliminate the redundant infleunce by its identifying mechnism, we normalize the AIF produced by different algorithms based on the DRI algorithm, denoted as $AIF^*$, where $AIF^*_X = (AIF_X - AIF_{DRI})/AIF_{DRI}$ and the variable $X$ repensents different algorithms. The Figue 4 has represented the redundant infleunce RI and normalized average infected fraction $AIF^*$ with the increasing number of spreaders $m$ from 10 to 100 based on a specifically infected probability $\beta$. From the Fig.4, one can find that the emergence of redundant influence is often accompanied by a obviously decline in the normalized average infected fration AIF*, for example, the AIF* of the NC algorithm decreases until have more spreaders than the DRI algorithm in all three networks. The spreaders identified by the DSN algorithm produce largest AIF* conpared with other baseline algorithms for continuously keeping seldom redundant infleunce when increasing the spreader number. Conversly, though both the CI and ND algorithms have tried to reduce the influence overlap of spreaders, they failed to completely eliminate the redundant influence, which result an inferior performance compared with the DRI and DSN algorithm on the Email and Soc-hamsterster networks. The results with other infected probability $\beta$ have been represented in Appedix.

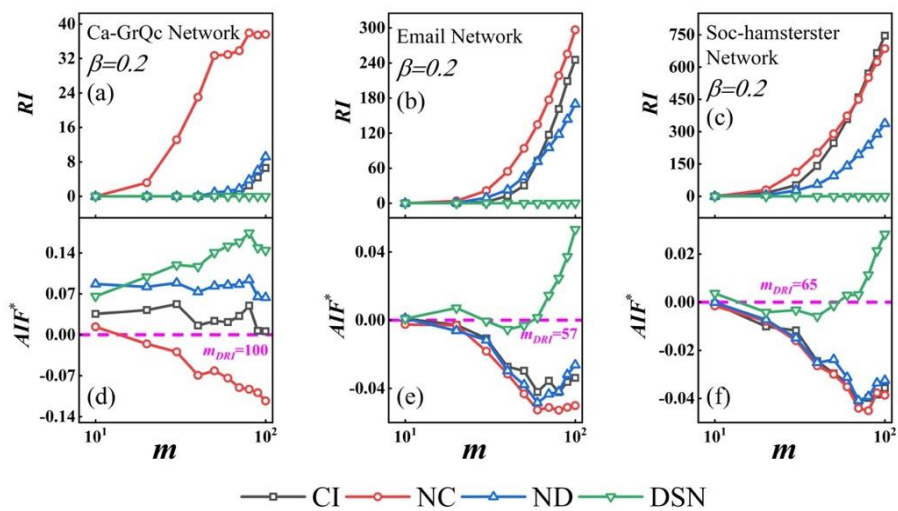

Figure 4: (Color online) The Redundant Influence RI (a)-(c) and the normalized Average Infected Fraction AIF* (d)-(f) with the increasing number of spreaders ($m$) identified by different algorithms. The normalized average infected fraction AIF* of different algorithms are normalized based on DRI algorithm (blue dot line), where AIF*$_X$ = (AIF$_X$ - AIF$_{DRI}$)/ AIF$_{DRI}$, and the varible $X$ represents different algorithm.

## 5 Practical properties of DSN algorithm

The simulation results represented in Fig.3 and Fig.4 have demonstrated the validity of our method for quantifying influence ovelarp and redundant infleunce. So in this sectoin we further test the practical properties of our algorithm. As the DRI algorithm has high compute complexity, we only consider about the DSN algorithm to test its practical properties from two aspects: stability test and spreader properties analysis. Previous work have found that it is difficult to get the complete structure information in real-world social networks (de la Haye et al., 2017; Jorgensen et al., 2018). So keeping stability with incompleted network information is important for multiple spreaders identification algorithm. Another vital question is identifying the spreaders advertised and cheap activated, which is defined as ecominic value (Growiec et al., 2018). This is why many companies now choose a group of web bloggers to recommend their products instead of celebrity endorsement (Morales et al., 2014).

5.1 Stability Test

As it is difficult to contain the whole network structure information in real-world social networks, we test the stability of the DSN algorithm by identifying the multiple spreaders on the network without part of nodes and their links ranging from 5% to 30%. The network may be disconnecting after removing part of nodes and edges, we therefore identify the spreaders from the largest connected components and implement simulations on the original network. We compare the average infected fraction AIF produced by the spreaders with different pair of spreader number $m$ and infected probability $\beta$, which are shown in the Figure 5. We can find from the Fig.5 that, for the same spreader number $m$ and infected probability $\beta$, the AIF produced by the spreaders identified from different networks without 5% to 30% structure information are similar with each other, and the largest variation is within 1% shown in the subgraph b of Fig.5, which suggests that the DSN algorithm is stable on the social network without complete structure information.

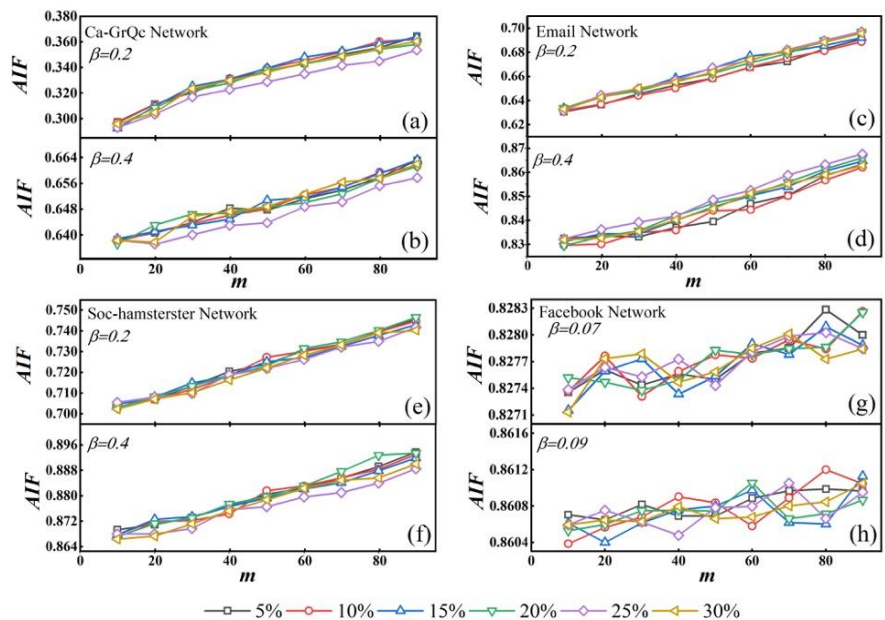

Figure 5: (Color online) The average infected fraction AIF produced by the spreaders for different networks and parameters.

5.2 The properties of spreaders

In most situation, we not only consider about the maximum spreading influence, but also care about the cost we spend on activating these spreaders. For example, in viral marketing, it is important for every marketing manager to select the users with low activation costs but can achieve good publicity to promote their products, which corresponding to the nodes with lower degree and $k$-core value (Venkatesh et al., 2016). Therefore, we compare the average degrees $\langle k \rangle$ and average $k$-core $\langle k\text{-core} \rangle$ of the spreaders selected by the DSN algorithm and other algorithms including the *CI, NC,* and *ND* algorithms. The results are showed in Fig.6, which suggests that the spreaders selected by the DSN algorithms have the smallest average degree $\langle k \rangle$ and average k-core $\langle k\text{-core} \rangle$ for most networks, which is more significant when increasing the number of spreaders.

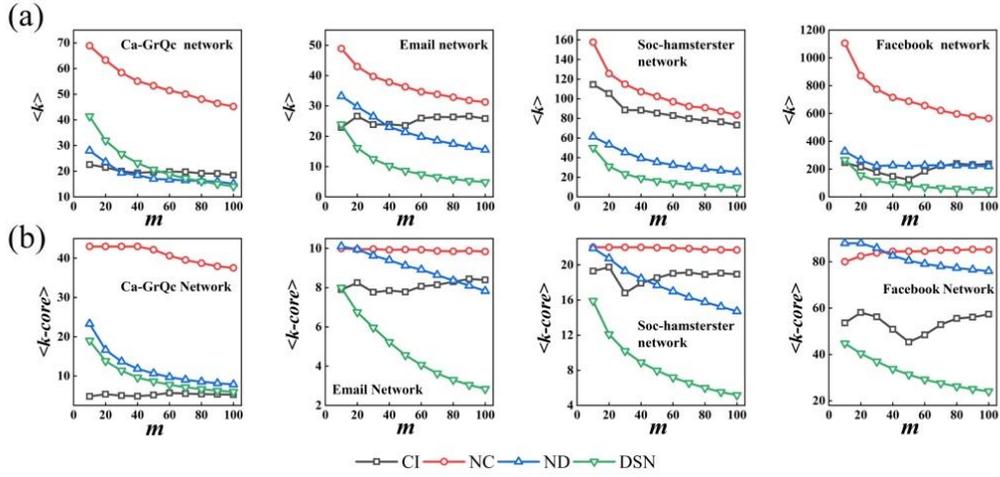

Figure 6: (Color online) The average degree ⟨$k$⟩ (a) and average $k$-core ⟨$k$-core⟩ (b) of the spreaders identified by four different algorithms on each social network.

## 6. Conclusions and future work

In this paper, we take a nuanced look at influence overlap and argue that the redundant influence generated by excessive influence overlap will cause negative effect for maximizing spreading influence. We first give a definition for the influence overlap, and then propose a theoretical method to quantify the influence overlap from the view of probability, that is, calculating the probability the non-spreader node will be infected by the spreader set. We define the Redundant Influence as the part of probability exceed the upper limit of a non-spreader node can be infected. Inspired by the idea of eliminating the redundant influence to maximize spreading influence, we present two algorithms for multiple spreaders identification. The first algorithm, namely Degree-Redundant-Influence algorithm, identifies the spreader in the order of node degree and keeping none redundant influence of non-spreader nodes by repeatedly traversing the whole network. The second algorithm, namely Degree-Second-order-Neighbors algorithm, identifies the spreaders in the order of node degree and eliminate the redundant influence by keeping the distance among each pair of spreaders exceed two, which is acquired by the solution of an optimal question focus on spreaders distribution.

We simulate the influence spreading with the SIR model for four empirical social networks, the experimental results compared with three state-in-the-art algorithms including the CI, NC and ND algorithms have demonstrated the validity of our theoretical method for measure the redundant influence. One can find that the spreaders generating more redundant influence produce less average infected fraction under the same infected probability. In addition, we find that the DRI algorithm can roughly pre-determine the necessary number of spreaders by automatically converging to the largest spreader set without redundant influence. And the DSN algorithm can identify the spreaders producing largest AIF in simulation, so we further test its practical properties by stability test and spreader properties analysis. The results show that the DSN algorithm is valid even 30% of the network structure information is missing, and the spreaders identified by the DSN algorithm have smaller average node degree and $k$-core values.

Though this work has contributed to the question of maximizing spreading influence from both theoretical and practical aspects, there are two aspects that deserve future consideration. An important and strong assumption in this paper is that all the nodes in the network are homogeneous susceptible to the influence spreading, while most nodes in real world prone to be heterogeneous . Recent research have demonstrated that the susceptibility of the nodes will be decided by their topological features (Sewell, 2018, 2017) and individual attributes (Larsen and Ellersgaard, 2017). One approach to combine the node heterogeneous information in our method is to create a function to measure infected probability, which could be designed to contain the factors of nodes topological features and individual attributes. While the current research have not produced a paradigm for measuring node heterogeneity, so the validity of this approach still need further investigation.

Another study worthy of further discussion is how to achieve maximum influence within a limited period. In the experiments of this paper, we find that eliminating the redundant influence has a significant improvement in extending the maximum spreading influence, but at the expense of the initial propagation speed. While the development of social platforms gives users the opportunity to choose more topics, so the duration of a topic that can become a hotspot and spread widely is not very long. So how to identify the spreaders can produce the maximizing spreading influence while taking into account the speed of information dissemination, it is worth more research.

**Acknowledgement**

This work is partially supported by the National Natural Science Foundation of China [Grant No. 61773248], the National Social Science Foundation of China [18ZDA088] and S-Tech Academic Support Program 2018——Internet Communication for Young Scholars, and the authors wish to thank Yang Liu for her helpful suggestions and comments.

**Reference**

Rossi, R.A., Ahmed, N.K., 2015a. The network data repository with interactive graph analytics and visualization, in: Twenty-Ninth AAAI Conference on Artificial Intelligence, pp. 4292–4293.

Hamsterster, . Hamsterster social network. Http://www.hamsterster.com.

Bae, J., Kim, S., 2014. Identifying and ranking influential spreaders in complex networks by neighborhood coreness. Phys. A Stat. Mech. its Appl. 395, 549–559. https://doi.org/10.1016/J.PHYSA.2013.10.047

Bozzo, E., Franceschet, M., 2013. Resistance distance, closeness, and betweenness. Soc. Networks 35, 460–469. https://doi.org/10.1016/j.socnet.2013.05.003

Burt, R.S., Talmud, I., 1993. Market niche. Soc. Networks 15, 133–149. https://doi.org/10.1016/0378-8733(93)90002-3

Cao, T., Wu, X., Wang, S., Hu, X., 2011. Maximizing influence spread in modular social networks by optimal resource allocation. Expert Syst. Appl. 38, 13128–13135. https://doi.org/10.1016/j.eswa.2011.04.119

Chang, B., Xu, T., Liu, Q., Chen, E.-H., 2018. Study on Information Diffusion Analysis in Social Networks and Its Applications. Int. J. Autom. Comput. 15, 377–401. https://doi.org/10.1007/s11633-018-1124-0

Chen, W., Wang, Y., 2009. Efficient Influence Maximization in Social Networks Categories and Subject Descriptors. Proc. 15th ACM SIGKDD Int. Conf. Knowl. Discov. data Min. 199–207. https://doi.org/10.1145/1557019.1557047

Christakis, N.A., Fowler, J.H., 2007. The Spread of Obesity in a Large Social Network over 32 Years. N. Engl. J. Med. 357, 370–379. https://doi.org/10.1056/NEJMsa066082

de la Haye, K., Embree, J., Punkay, M., Espelage, D.L., Tucker, J.S., Green, H.D., 2017. Analytic strategies for longitudinal networks with missing data. Soc. Networks 50, 17–25.

https://doi.org/10.1016/J.SOCNET.2017.02.001

De Stefano, D., Fuccella, V., Vitale, M.P., Zaccarin, S., 2013. The use of different data sources in the analysis of co-authorship networks and scientific performance. Soc. Networks 35, 370–381. https://doi.org/10.1016/J.SOCNET.2013.04.004

Ertem, Z., Veremyev, A., Butenko, S., 2016. Detecting large cohesive subgroups with high clustering coefficients in social networks. Soc. Networks 46, 1–10. https://doi.org/10.1016/J.SOCNET.2016.01.001

Fang, X., Hu, P.J.-H., 2018. Top Persuader Prediction for Social Networks. MIS Q. 42, 63–82. https://doi.org/10.25300/MISQ/2018/13211

Flores, R., Koster, M., Lindner, I., Molina, E., 2012. Networks and collective action. Soc. Networks 34, 570–584. https://doi.org/10.1016/J.SOCNET.2012.06.003

Granovetter, M.S., 1977. The Strength of Weak ties. Soc. Networks. Acdemic Press 347–367. https://doi.org/10.1086/225469

Growiec, K., Growiec, J., Kamiński, B., 2018. Social network structure and the trade-off between social utility and economic performance. Soc. Networks 55, 31–46. https://doi.org/10.1016/J.SOCNET.2018.05.002

Guo, L., Lin, J.H., Guo, Q., Liu, J.G., 2016. Identifying multiple influential spreaders in term of the distance-based coloring. Phys. Lett. Sect. A Gen. At. Solid State Phys. 380, 837–842. https://doi.org/10.1016/j.physleta.2015.12.031

Harrigan, N., Achananuparp, P., Lim, E.-P., 2012. Influentials, novelty, and social contagion: The viral power of average friends, close communities, and old news. Soc. Networks 34, 470–480. https://doi.org/10.1016/J.SOCNET.2012.02.005

Hu, Y., Ji, S., Jin, Y., Feng, L., Stanley, H.E., Havlin, S., 2018. Local structure can identify and quantify influential global spreaders in large scale social networks. Proc. Natl. Acad. Sci. 115, 7468–7472. https://doi.org/10.1073/pnas.1710547115

Hu, Z.L., Liu, J.G., Yang, G.Y., Ren, Z.M., 2014. Effects of the distance among multiple spreaders on the spreading. Epl 106, 6–9. https://doi.org/10.1209/0295-5075/106/18002

Iribarren, J.L., Moro, E., 2011. Affinity Paths and information diffusion in social networks. Soc. Networks 33, 134–142. https://doi.org/10.1016/j.socnet.2010.11.003

Johnson, R., Kovács, B., Vicsek, A., 2012. A comparison of email networks and off-line social networks: A study of a medium-sized bank. Soc. Networks 34, 462–469. https://doi.org/10.1016/J.SOCNET.2012.02.004

Jorgensen, T.D., Forney, K.J., Hall, J.A., Giles, S.M., 2018. Using modern methods for missing data analysis with the social relations model: A bridge to social network analysis. Soc. Networks 54, 26–40. https://doi.org/10.1016/J.SOCNET.2017.11.002

Jung, K., Heo, W., Chen, W., 2012. IRIE: Scalable and Robust Influence Maximization in Social Networks, in: 2012 IEEE 12th International Conference on Data Mining. IEEE, pp. 918–923. https://doi.org/10.1109/ICDM.2012.79

Kempe, D., Kleinberg, J., Tardos, É., 2003. Maximizing the spread of influence through a social network. Proc. ninth ACM SIGKDD Int. Conf. Knowl. Discov. data Min. - KDD '03 137. https://doi.org/10.1145/956755.956769

Kitsak, M., Gallos, L.K., Havlin, S., Liljeros, F., Muchnik, L., Stanley, H.E., Makse, H.A., 2010. Identification of influential spreaders in complex networks. Nat. Phys. 6, 888–893. https://doi.org/10.1038/nphys1746

Korn, F., Muthukrishnan, S., 2000. Influence sets based on reverse nearest neighbor queries, in: Proceedings of the 2000 ACM SIGMOD International Conference on Management of Data - SIGMOD '00. ACM Press, New York, New York, USA, pp. 201–212. https://doi.org/10.1145/342009.335415

Larsen, A.G., Ellersgaard, C.H., 2017. Identifying power elites—k-cores in heterogeneous affiliation networks. Soc. Networks 50, 55–69. https://doi.org/10.1016/j.socnet.2017.03.009

Lin, J.-H.H., Guo, Q., Dong, W.-Z.Z., Tang, L.-Y.Y., Liu, J.-G.G., 2014. Identifying the node spreading influence with largest k-core values. Phys. Lett. Sect. A Gen. At. Solid State Phys. 378, 3279–3284. https://doi.org/10.1016/j.physleta.2014.09.054

Liu, H.L., Ma, C., Xiang, B.B., Tang, M., Zhang, H.F., 2018. Identifying multiple influential spreaders based on generalized closeness centrality. Phys. A Stat. Mech. its Appl. 492, 2237–2248. https://doi.org/10.1016/j.physa.2017.11.138

Liu, J.-G., Wang, Z.-Y., Guo, Q., Guo, L., Chen, Q., Ni, Y.-Z., 2017. Identifying multiple influential spreaders via local structural similarity. EPL (Europhysics Lett. 119, 18001. https://doi.org/10.1209/0295-5075/119/18001

Liu, Y., Tang, M., Zhou, T., Do, Y., 2016. Identify influential spreaders in complex networks, the role of neighborhood. Phys. A Stat. Mech. its Appl. 452, 289–298. https://doi.org/10.1016/j.physa.2016.02.028

Lopaciuk-Gonczaryk, B., 2016. Collaboration strategies for publishing articles in international journals – A study of Polish scientists in economics. Soc. Networks 44, 50–63. https://doi.org/10.1016/J.SOCNET.2015.07.001

Lü, L., Chen, D., Ren, X.-L., Zhang, Q.-M., Zhang, Y.-C., Zhou, T., 2016a. Vital nodes identification in complex networks. Phys. Rep. 650, 1–63. https://doi.org/10.1016/j.physrep.2016.06.007

Lü, L., Zhou, T., Zhang, Q.-M., Stanley, H.E., 2016b. The H-index of a network node and its relation to degree and coreness. Nat. Commun. 7, 10168. https://doi.org/10.1038/ncomms10168

Ma, L.L., Ma, C., Zhang, H.F., Wang, B.H., 2016. Identifying influential spreaders in complex networks based on gravity formula. Phys. A Stat. Mech. its Appl. 451, 205–212. https://doi.org/10.1016/j.physa.2015.12.162

Morales, A.J.J., Borondo, J., Losada, J.C.C., Benito, R.M.M., 2014. Efficiency of human activity on information spreading on Twitter. Soc. Networks 39, 1–11. https://doi.org/10.1016/j.socnet.2014.03.007

Morone, F., Makse, H.A., 2015. Influence maximization in complex networks through optimal percolation. Nature 524, 65–68. https://doi.org/10.1038/nature14604

Ott, M.Q., Light, J.M., Clark, M.A., Barnett, N.P., 2018. Strategic players for identifying optimal social network intervention subjects. Soc. Networks 55, 97–103. https://doi.org/10.1016/J.SOCNET.2018.05.004

Pastor-Satorras, R., Vespignani, A., 2002. Epidemic dynamics in finite size scale-free networks. Phys. Rev. E 65, 035108. https://doi.org/10.1103/PhysRevE.65.035108

Piedrahita, P., Borge-Holthoefer, J., Moreno, Y., González-Bailón, S., 2018. The contagion effects of repeated activation in social networks. Soc. Networks 54, 326–335. https://doi.org/10.1016/j.socnet.2017.11.001

Sewell, D.K., 2018. Heterogeneous susceptibilities in social influence models. Soc. Networks 52, 135–144. https://doi.org/10.1016/J.SOCNET.2017.06.004

Sewell, D.K., 2017. Network autocorrelation models with egocentric data. Soc. Networks 49, 113–123. https://doi.org/10.1016/J.SOCNET.2017.01.001

Shrestha, M., Scarpino, S. V., Moore, C., 2015. Message-passing approach for recurrent-state epidemic models on networks. Phys. Rev. E 92, 022821. https://doi.org/10.1103/PhysRevE.92.022821

Sohn, M.-W., 2001. Distance and cosine measures of niche overlap. Soc. Networks 23, 141–165.

https://doi.org/10.1016/S0378-8733(01)00039-9

Valente, T.W., Fujimoto, K., 2010. Bridging: Locating critical connectors in a network. Soc. Networks 32, 212–220. https://doi.org/10.1016/J.SOCNET.2010.03.003

Venkatesh, V., Rai, A., Sykes, T.A., Aljafari, R., 2016. Combating Infant Mortality in Rural India: Evidence from a Field Study of eHealth Kiosk Implementations. MIS Q. 40, 353–380. https://doi.org/10.25300/MISQ/2016/40.2.04

Ventresca, M., Aleman, D., 2013. Evaluation of strategies to mitigate contagion spread using social network characteristics. Soc. Networks 35, 75–88. https://doi.org/10.1016/j.socnet.2013.01.002

Vriens, E., Corten, R., 2018. Are bridging ties really advantageous? An experimental test of their advantage in a competitive social learning context. Soc. Networks 54, 91–100. https://doi.org/10.1016/J.SOCNET.2018.01.007

Wang, S., Du, Y., Deng, Y., 2017. A new measure of identifying influential nodes: Efficiency centrality. Commun. Nonlinear Sci. Numer. Simul. 47, 151–163. https://doi.org/10.1016/j.cnsns.2016.11.008

Wood, G., 2017. The structure and vulnerability of a drug trafficking collaboration network. Soc. Networks 48, 1–9. https://doi.org/10.1016/j.socnet.2016.07.001

Xia, Y., Ren, X., Peng, Z., Zhang, J., She, L., 2016. Effectively identifying the influential spreaders in large-scale social networks. Multimed. Tools Appl. 75, 8829–8841. https://doi.org/10.1007/s11042-014-2256-z

Yamaguchi, K., 2002. The structural and behavioral characteristics of the smallest-world phenomenon: minimum distance networks. Soc. Networks 24, 161–182. https://doi.org/10.1016/S0378-8733(01)00055-7

Yang, X., Huang, D., 2017. Maximize Influence Diffusion in Social Networks with Spreading Distance and 2-Step Neighborhood Overlapping Effects. J. Internet Technol. 18, 653–665. https://doi.org/10.6138/JIT.2017.18.3.20161203

Zhang, J.X., Chen, D.B., Dong, Q., Zhao, Z.D., 2016. Identifying a set of influential spreaders in complex networks. Sci. Rep. 6, 1–10. https://doi.org/10.1038/srep27823

Zhang, Z.-K., Liu, C., Zhan, X.-X., Lu, X., Zhang, C.-X., Zhang, Y.-C., 2016. Dynamics of information diffusion and its applications on complex networks. Phys. Rep. 651, 1–34. https://doi.org/10.1016/j.physrep.2016.07.002

Zhao, X.Y., Huang, B., Tang, M., Zhang, H.F., Chen, D.B., 2014. Identifying effective multiple spreaders by coloring complex networks. EPL 108, 1–6. https://doi.org/10.1209/0295-5075/108/68005

Zhou, M.Y., Xiong, W.M., Wu, X.Y., Zhang, Y.X., Liao, H., 2018. Overlapping influence inspires the selection of multiple spreaders in complex networks. Phys. A Stat. Mech. its Appl. 508, 76–83. https://doi.org/10.1016/j.physa.2018.05.022

Zhou, S., Cox, I.J., Hansen, L.K., 2017. Second-Order Assortative Mixing in Social Networks, in: International Workshop on Complex Networks. Springer, Cham. pp. 3–15.